\documentclass[twocolumn,showpacs,preprintnumbers,amsmath,amssymb]{revtex4}
\pdfoutput=1
\usepackage{multirow}
\usepackage{graphicx}% Include figure files
\usepackage{dcolumn}% Align table columns on decimal point
\usepackage{bm}% bold math
\newcommand{\micron}{$\mu$m}
\newcommand{\microns}{$\mu$m }
\renewcommand{\deg}{$^\circ$}

\begin{document}

%\preprint{APS/123-QED}

\title{Split Cylinder Resonators with a New Magnetic Resonance\\ in the Midinfrared under Normal Incidence }

\author{Sher-Yi Chiam}
 \email{phycsy@nus.edu.sg}
\author{Andrew A. Bettiol}
\author{JiaGuang Han}
\author{Frank Watt}
\affiliation
{Department of Physics, Science Drive 3, National University of Singapore, Singapore 117542}

\author{Mohammed Bahou}
\author{Herbert O. Moser}
\affiliation
{Singapore Synchrotron Light Source, 5 Research Link, National University of Singapore, Singapore 117603}

\date{\today}% It is always \today, today,
             %  but any date may be explicitly specified

\begin{abstract}
So far, research in the field of metamaterials has been carried out largely with arrays of flat, 2-dimensional structures. Here, we report a newly identified magnetic resonance in Split Cylinder Resonators (SCRs), a 3-dimensional version of the Split Ring Resonator (SRR), which were fabricated with the Proton Beam Writing technique.  Experimental and numerical results indicate a hitherto unobserved 3-dimensional resonance mode under normal incidence at about 26 THz, when the SCR depth is approximately half the free space wavelength. This mode is characterized by strong currents along the cylinder axis which are concentrated at the cylinder gaps. Due to their orientation, these axial currents give rise to a magnetic response under normal incidence, which is not possible in shallow SRRs. Our results reveal new behavior in the SRR structure which arises from a change in its aspect ratio. Such new resonances can have a significant influence on the quest for practical, 3-dimensional metamaterials.

\end{abstract}

% PACS, the Physics and Astronomy
                             % Classification Scheme.
%Use showkeys class option if keyword
                              %display desired
\maketitle

Metamaterials are artificial composites which enable magnetism to be achieved over a range of frequencies \cite{pendrysrr,yen,linden}. There has been a great deal of research on metamaterials in recent years. The split ring resonator (SRR) first proposed by Pendry \emph{et al} \cite{pendrysrr} has played a pioneering role in this research.  SRRs consist of two concentric metallic rings with gaps situated oppositely. This design allows resonances where inductive currents circulate along the rings in conjunction with capacitive charge accumulation at the gaps.  These circular currents, when excited by an external oscillating magnetic field, result in a magnetic response and thus considerably influence the effective permeability ($\mu$) of the material. This can result in negative effective $\mu$ over a frequency range close to the resonance \cite{smith,pendrysrr}. The circular currents can also be excited by an oscillating electric field parallel to the gap sides of the SRR \cite{gay,katsa_elect}. We shall refer to resonances with circular currents as LC resonances.  SRRs are also shown to have an electrical response similar to that of cut wires\cite{koschny}. These electrical resonances are due to antenna-like couplings between the SRRs and the incident electric field and can result in a region of negative effective electric permittivity ($\epsilon$).

To date, SRRs have been experimentally studied over a wide frequency range, from the low GHz \cite{smith,shelby}, to
the THz regime \cite{moserprl,padilla_prl,yen}, and finally to the near infrared \cite{linden,enkrichprl}. Obtaining a magnetic response from SRRs requires the presence of significant magnetic field normal to the SRR plane. Thus the incident radiation must propagate along (or obliquely) to the SRR plane. The lack of a magnetic response from SRRs under normal incidence has been a stumbling block its use in the construction of practical materials. For practical applications, planar arrays of SRRs need to be stacked to give greater width perpendicular to the propagation direction. For example, Shelby \emph{et al} stacked printed circuit boards to fabricate a prism used to demonstrate negative refraction in the Gigahertz range \cite{shelby}. At higher frequencies, where nanofabrication techniques are needed, stacking becomes challenging. However, some effort has been made in this direction. Katsarakis \emph{et al} fabricated a metamaterial of 5 layers of single split rings (SSRs) resonating in the far infrared regime ($\sim6$ THz) \cite{katsarakis}. Liu \emph{et al} reported stacking 4 or more layers of sub-micron SSRs operating at about 100 THz \cite{liu3D}. The processes used in these works resulted in split ring structures separated by dielectric spacers.

Some authors have applied 3-dimensional lithography techniques to fabricate nanostructures capable of coupling with the external magnetic field under normal incidence. For example, Zhang \textit{et al} have used a process based on interference lithography which resulted in Au ``staples" deposited on a pitch grating\cite{staples}. Removal of the pitch grating resulted in Au staples standing upright on the substrate. These structures exhibited a magnetic response under normal incidence when a the incoming magnetic field is perpendicular to the plane of the staples. Very recently, Rill \emph{et al} fabricated a planar structure consisting of connected, elongated SRRs as a starting point for a stacked, ``woodpile" structure of elongated SRRs \cite{rill}.

Another possible alternative approach is to fabricate, in a single lithography step, high aspect ratio structures with great depth perpendicular to the lithography plane. Recently, Casse \emph{et al} have used deep X-ray lithography for this purpose, demonstrating its application for resist 200 \microns thick \cite{moseapl}. Such an approach would avoid some of the difficulties, such as alignment issues, arising from layering techniques. Furthermore deep structures allow currents to flow vertically (perpendicular to the lithography plane). This can result in distinct resonant modes unavailable in stacked, 2-dimensional structures separated by spacers.

Here, we report on the the fabrication of  single layers of very deep Split Ring Resonators (SRRs) using a focused, sub-micron MeV proton beam. These Split Cylinder Resonators (SCRs) have excellent sidewall quality and high aspect ratio. We also present results from spectral measurements made using Fourier Transform Infra-Red (FTIR) Spectroscopy as well as simulated results obtained using the commercially available Microwave Studio\texttrademark\space software. Our experimental and numerical results give evidence of a hitherto unobserved magnetic resonance ($\sim$26 THz) under normal incidence, which is not possible with convectional SRRs.

A number of previous works have studied the effect of SRR depth on their LC resonance \cite{okthickness,myspie,guo}, although not at the high aspect ratios of this current work. Here, we fabricated and characterized 2 SCR samples with depths in excess of their ring diameters. For comparison, we also fabricated a regular SRR sample of lower depth,  as well as closed rings and closed cylinders (where the gaps in the SRRs are eliminated).

\begin{figure}[h!]
    \begin{center}
    \begin{tabular}{cc}
   \includegraphics[height=4cm]{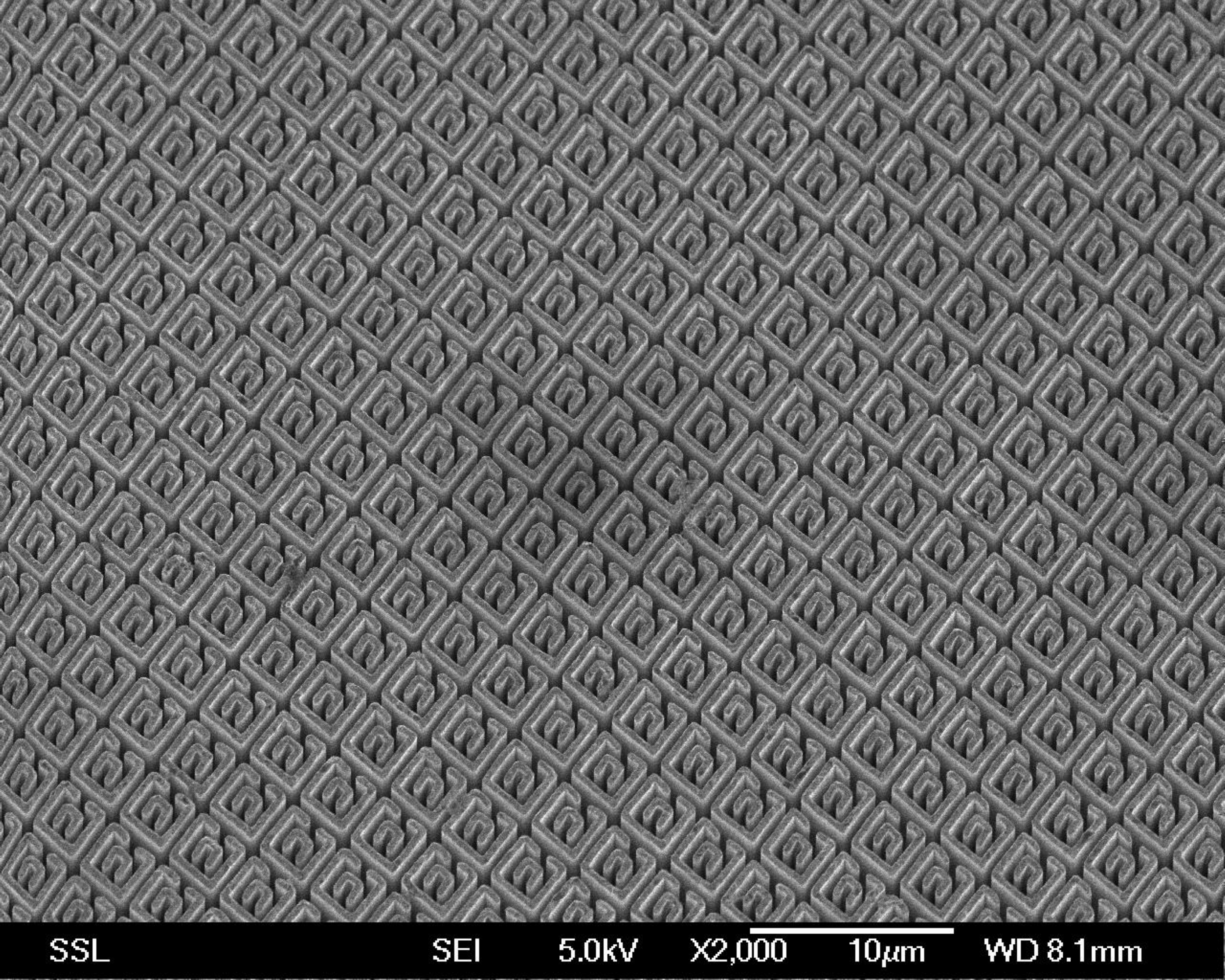}& \multirow{4}{*}{\includegraphics[height=2cm]{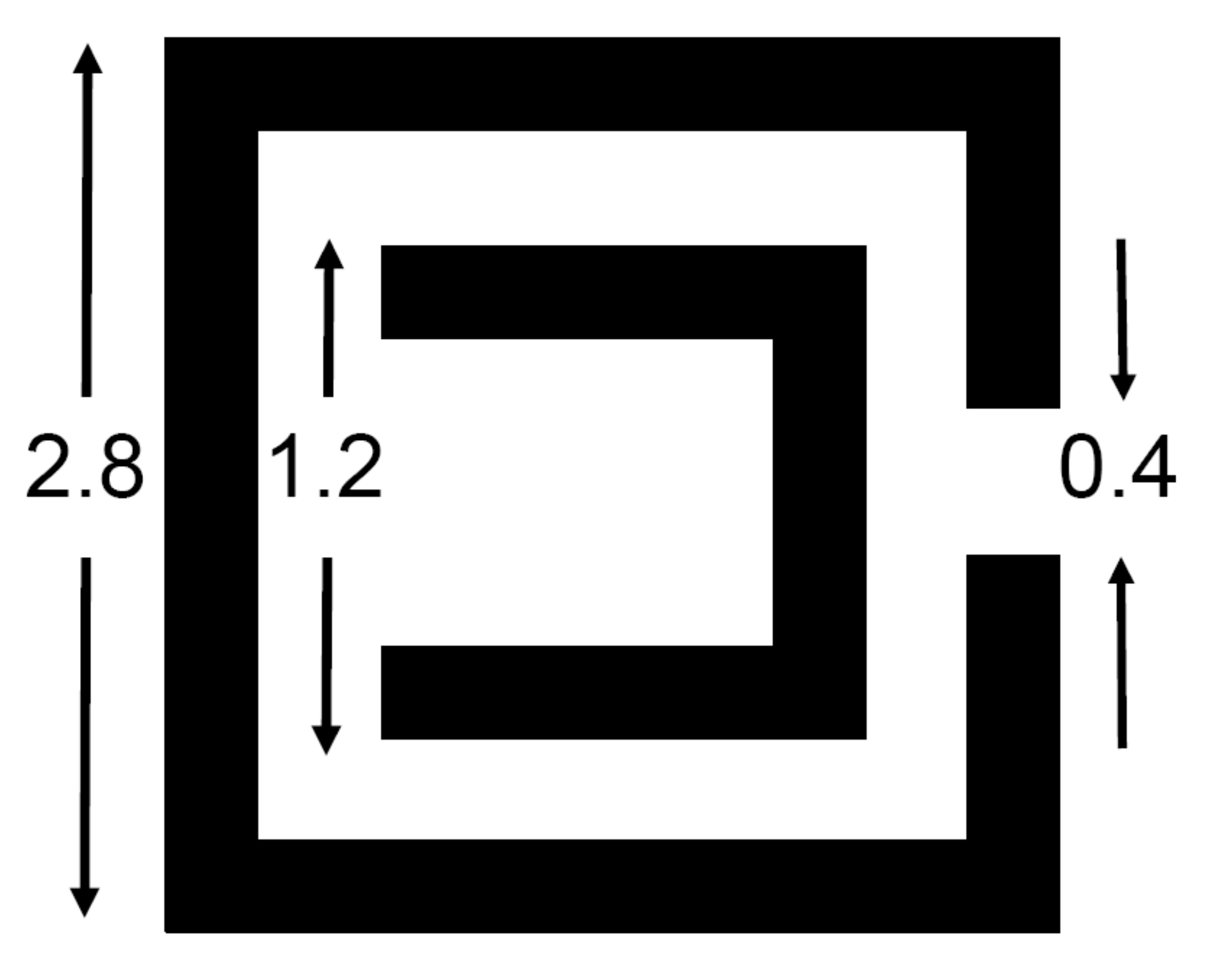}}\\ \bf{a}&\\
   \includegraphics[height=4cm]{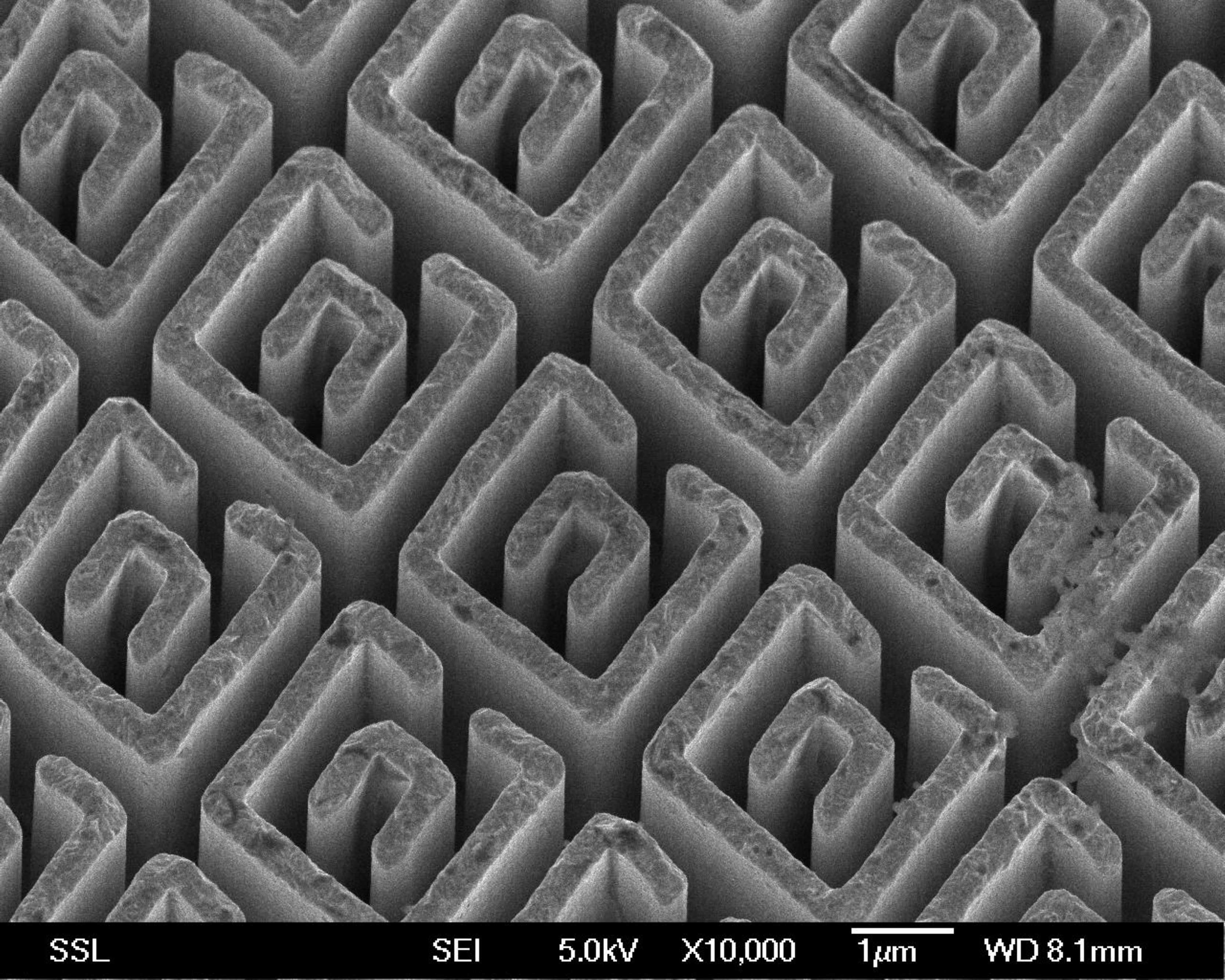}& \\
   \bf{b}&\\
   \end{tabular}
   \end{center}
   \vspace{-0.5cm}
   \caption[srrsem]
{\label{srrsem}
 Scanning electron micrographs of gold SRR arrays fabricated for this work. Sample quality is high over a large area (\textbf{a}), and individual structures have good sidewall quality (\textbf{b}). The depth of the SRRs shown is about 5.6 \micron. The designed lateral dimensions are given in microns in the insert. Linewidth is approximately 350nm and the unit cell distance is 3.2 \micron .}
\end{figure}

Fabrication work was carried out at the Center for Ion Beam Applications at the National University of Singapore using a fabrication process based on the direct write Proton Beam Writing (PBW) technique \cite{pbwapl,pbwmat2day}. Si wafers were first sputtered with thin ($\sim$20 nm) layers of chrome (Cr) and gold (Au), which served respectively as adhesion and electroplating seed layers. Proton Beam Writing was then used to write a latent image in polymethylmethacrylate (PMMA) resist spin coated onto the Si wafers.  PBW utilizes a highly focused Mega-electronvolt proton beam to write latent images in resist. As protons maintain relatively straight tracts through tens of microns of resist, the technique allows fabrication of deep structures with vertical sidewalls. After writing an image in PMMA resist of sufficient depth, we used an electroplating step to define SRR structures in gold, using the resist as an electroplating mould. This allows higher aspect ratios than evaporation or sputtering techniques. The depth of the SRRs in this case is determined by the plating current and time. Care was taken to avoid overplating. The resist is then chemically stripped after serving its function as a plating mould. To prevent shorting of the SRRs, the sputtered Au and Cr layers were chemically etched off. Au etching times were carefully controlled to prevent damage to the Au SRRs, and a highly specific chemical etch was used for the Cr layer. Individual SRR structures have critical dimensions of around 350 nm and depths up to 5 \microns (Figure \ref{srrsem}). The unit cell of each SRR is 3.2\micron, with each array measuring 200 \microns by 200 \micron. Each array thus contains over 4000 individual SRRs.

The samples were characterized using a Bruker Hyperion 2000 IR microscope coupled to a Bruker IFS 66v/S Fourier
Transform Infrared spectrometer (FTIR). Spectra at normal incidence were collected in reflection as well as transmission mode under different polarizations using the bare silicon substrate as reference. The beam spot covered almost the entire array of 4000 structures. A KBr beamsplitter and mid-band MCT infrared detector cooled to 77 K were used. The numerical aperture of the Schwarzschild infrared objective used was 0.4, corresponding to a maximum conical incidence angle of 23\deg .

\begin{figure*}[htbp]
    \begin{center}
  \includegraphics[height=9.5cm]{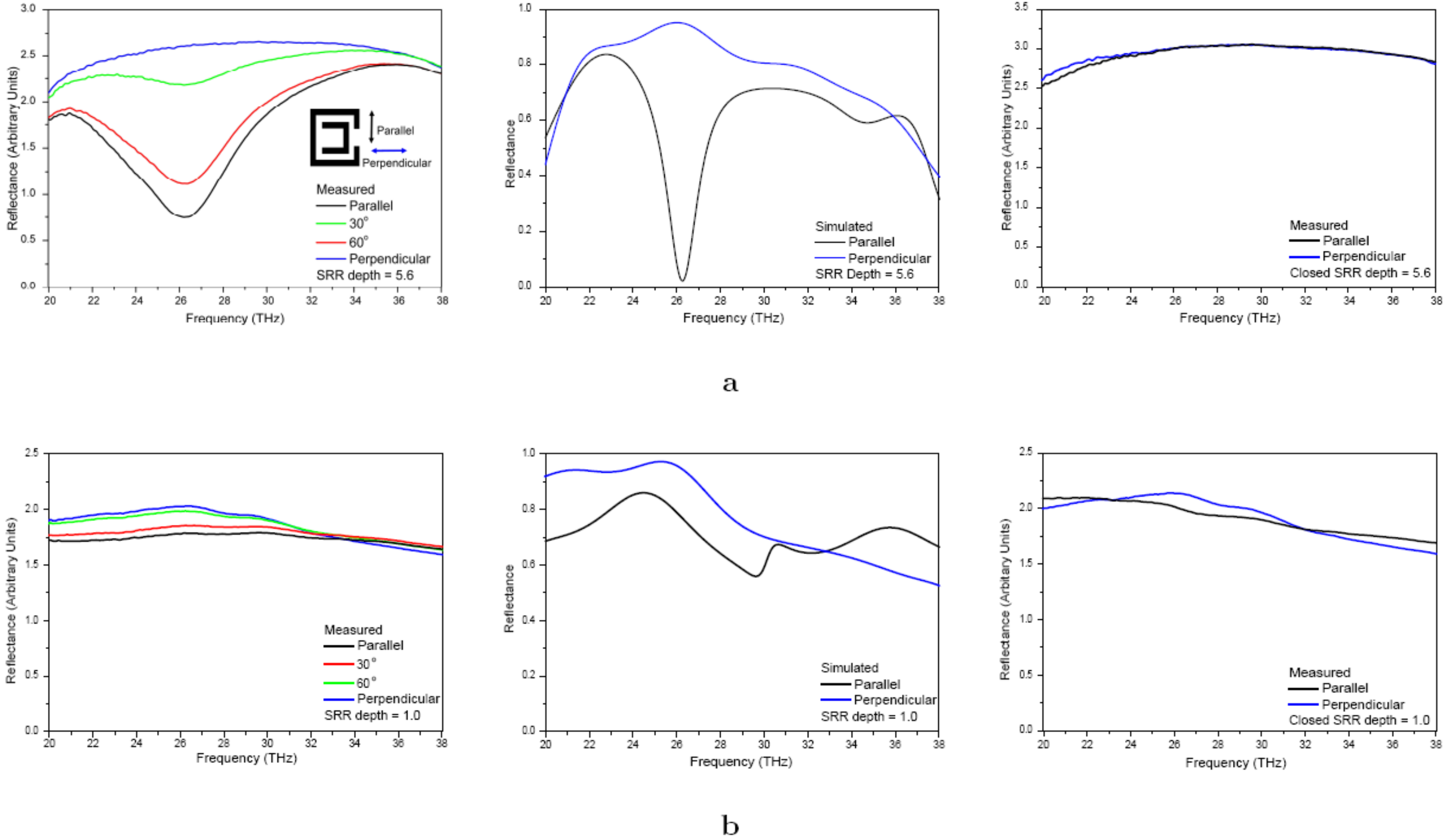}
    \end{center}
      \vspace{-0.5cm}
   \caption[spec]
{\label{spec} Measured (left panels) and simulated (center panels) reflection spectra of SRRs with depth of 5.6 \microns (\textbf{a}) and 1.0 \microns (\textbf{b}). Right panels show measured spectra for closed rings. Insert in (\textbf{a}) shows the orientation of the electric field relative to the SRRs for parallel and perpendicular polarization.}
\end{figure*}

Simulations were carried out using the commercial Microwave Studio\texttrademark\space software package. Due to the very tight packing of our arrays, which must lead to significant coupling between individual structures, we found that simulating a single unit cell resulted in spectra that were slightly red shifted relative to experimental results. Simulations were thus carried out for a four by four array of SRRs, where the top row of SRRs (i.e. those without gap-side neighbors) had their gaps closed. Closing the gaps of the top row destroyed their resonance and suppressed their red-shifted spectral contributions. The simulation domain had perfect electrical and perfect magnetic conductor boundary conditions for the sides and open boundary conditions for the ends. We modeled gold as a lossy metal with conductivity = $4.09 \times 10^7 Sm^{-1}$. The electrical permittivity of the silicon substrate is taken to be 11.6 \cite{grischkowsky} with tangent $\delta = 4 \times 10^{-3}$

The measured spectral response (with sample plane normal to beam axis) of two SRR samples and their closed ring versions are shown in Figure \ref{spec}. A prominent reflection dip is seen at about 26 THz for the 5.6 \microns deep SRR sample. This dip is strongest under parallel polarization, when the electric field is parallel to the gap sides of the SRRs. It weakens gradually at other polarizations and disappears completely under perpendicular polarization. For the 1 \microns deep SRRs, no reflection dip is present from 20 THz to 38 THz. For closed rings, the spectra are polarization independent, regardless of depth. Simulated data match the experimental data well.

\begin{figure}[h!]
    \begin{center}
    \begin{tabular}{c}
    \includegraphics[height=4cm]{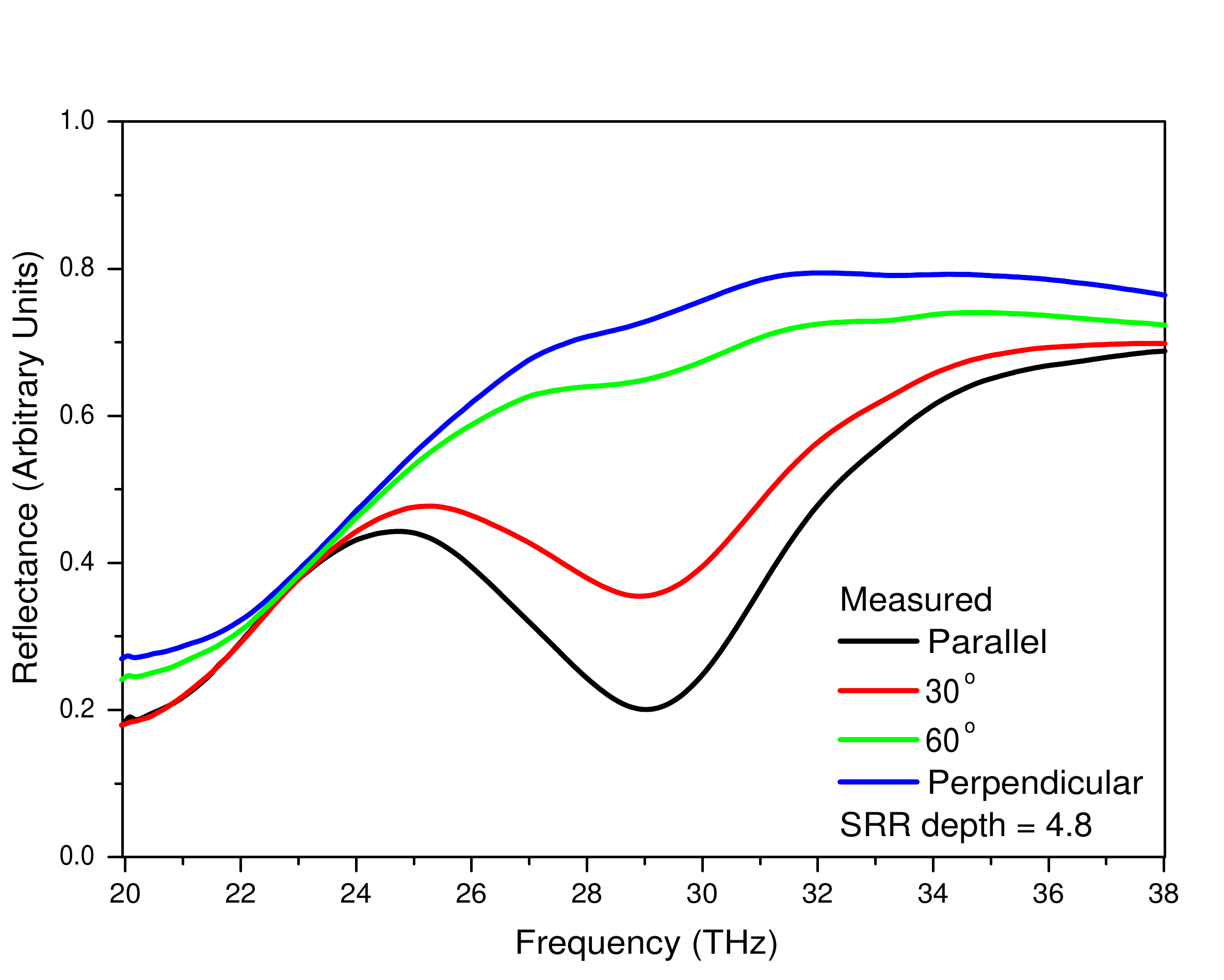}\\\bf{a}\\
    \includegraphics[height=4cm]{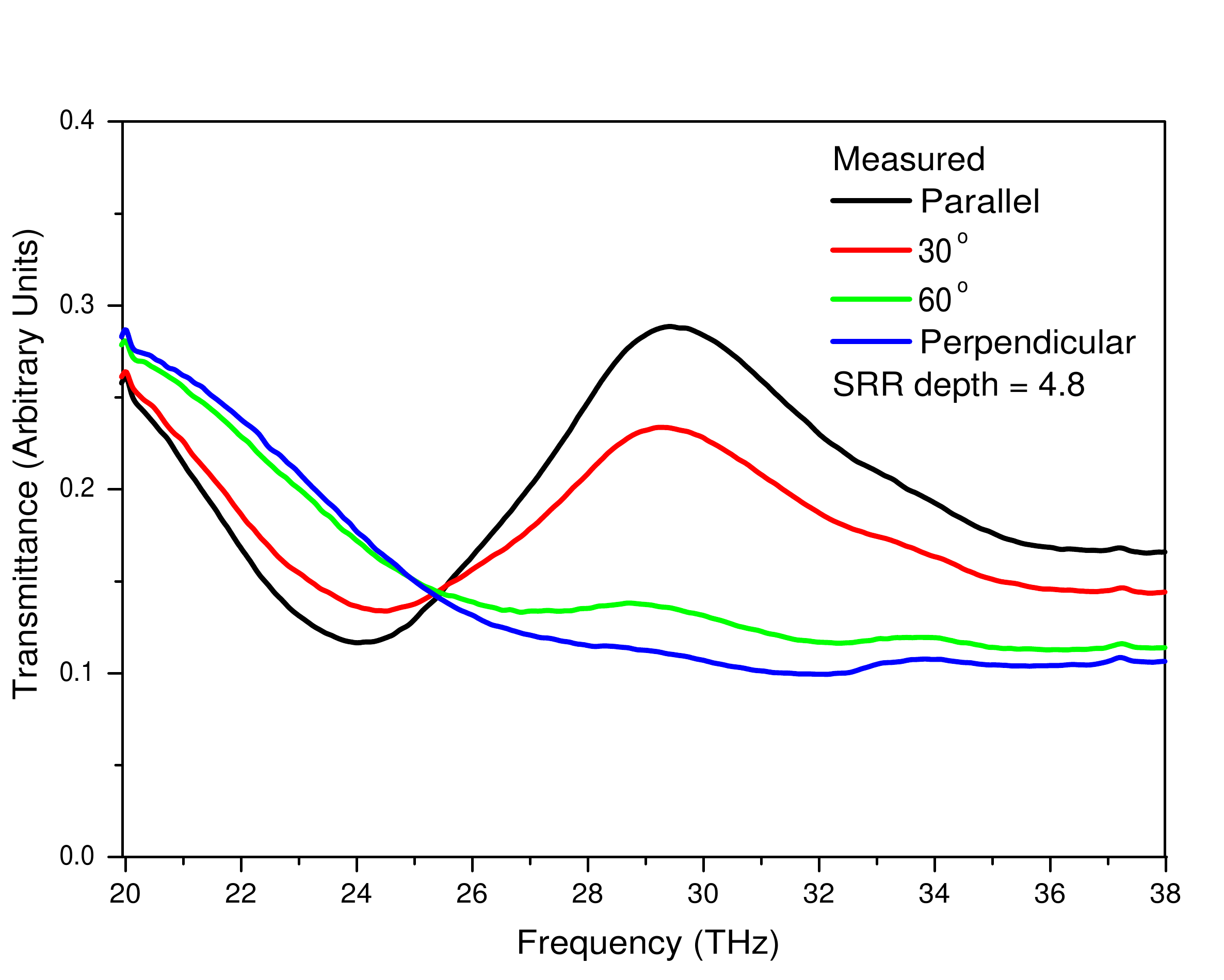}\\\bf{b}\\
   \end{tabular}
   \end{center}
   \vspace{-0.5cm}
   \caption[tspec]
{\label{tspec} Measured reflectance (\textbf{a}) and transmittance (\textbf{b}) for SRRs of height 4.8 \micron. The resonance occurs at a higher frequency than for the 5.6 \microns deep SRR. }
\end{figure}

The origin of the reflection dip in the deep SRRs is the main focus of this letter.  Being present only in SRRs under parallel polarization and absent in closed rings, it appears to be associated with an LC resonance. However, LC resonances in SRRs under normal incidence typically result in transmission dips, instead of reflection dips. Normal incidence results in there being no magnetic field normal to the rings. In shallow SRRs, the LC resonance can only couple to the external electric field, influencing solely the behavior of $\epsilon$ \cite{katsa_elect}.  A region of negative $\epsilon$ without negative $\mu$ leads to a stop band. To further investigate the nature of the reflection dip, we measured both the reflection and transmission (Figure \ref{tspec}) for an additional SRR sample with depth of 4.8 \micron .

This revealed that the reflection dip is accompanied by a corresponding transmission peak, clearly indicating the presence of a passband. We also observe that the resonant frequency shifts downwards with increasing SRR depth. This trend, which was captured in our simulations, is also unexpected.  Previous experimental studies have shown that the frequency of LC resonances shifts \emph{upward} with depth \cite{guo,myspie}. These observations indicate that the reflection dips are due to a resonance other than the regular LC resonance.

\begin{figure}[htbp]
\begin{center}
 \includegraphics[height=8cm]{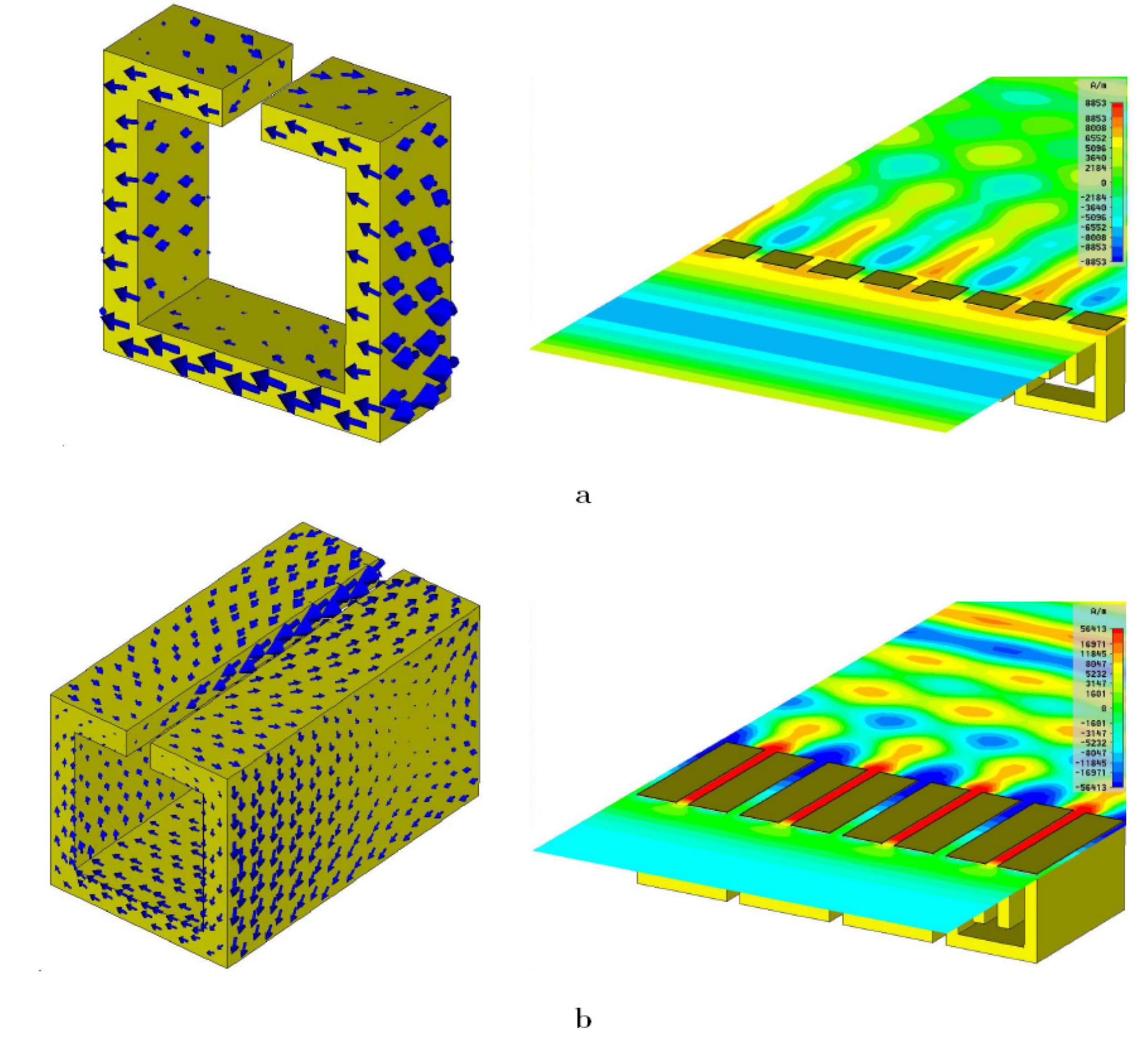}
\end{center}
  \vspace{-0.5cm}
\caption[current] {\label{current} Simulated surface currents (left panels) and magnetic fields (right panels) for SRRs
of depth 1.0\microns (\textbf{a}) and 5.6\microns (\textbf{b}). The magnetic field component plotted is the one normal to the planes shown. Note the different color scales - maximum field is several times larger in \textbf{b}.}
\end{figure}

Figure \ref{current} shows simulated surface current and magnetic field snapshots for SRRs of depths 1.0 \microns and 4.8 \microns under parallel polarization. The frequency for both snapshots is 27.3 THz, where the deeper SRR shows a reflection dip. For the 1.0 \microns SRR, the current does not show the characteristic circular pattern of LC resonances and is driven mainly by the electric field with no coupling to the external magnetic field. In the 4.8 \microns deep SRR, we observe circular currents along the cylinder circumference, as well as strong currents flowing along the cylinder axis. The axial currents, which are concentrated at the gaps, flow up on one side and down the other. This gives rise to enhanced gap fields and results in a magnetic response in deep SRRs. This feature is absent in the shallow SRRs.

\begin{figure}[htbp]
\begin{center}
\begin{tabular}{c}
\includegraphics[height=4cm]{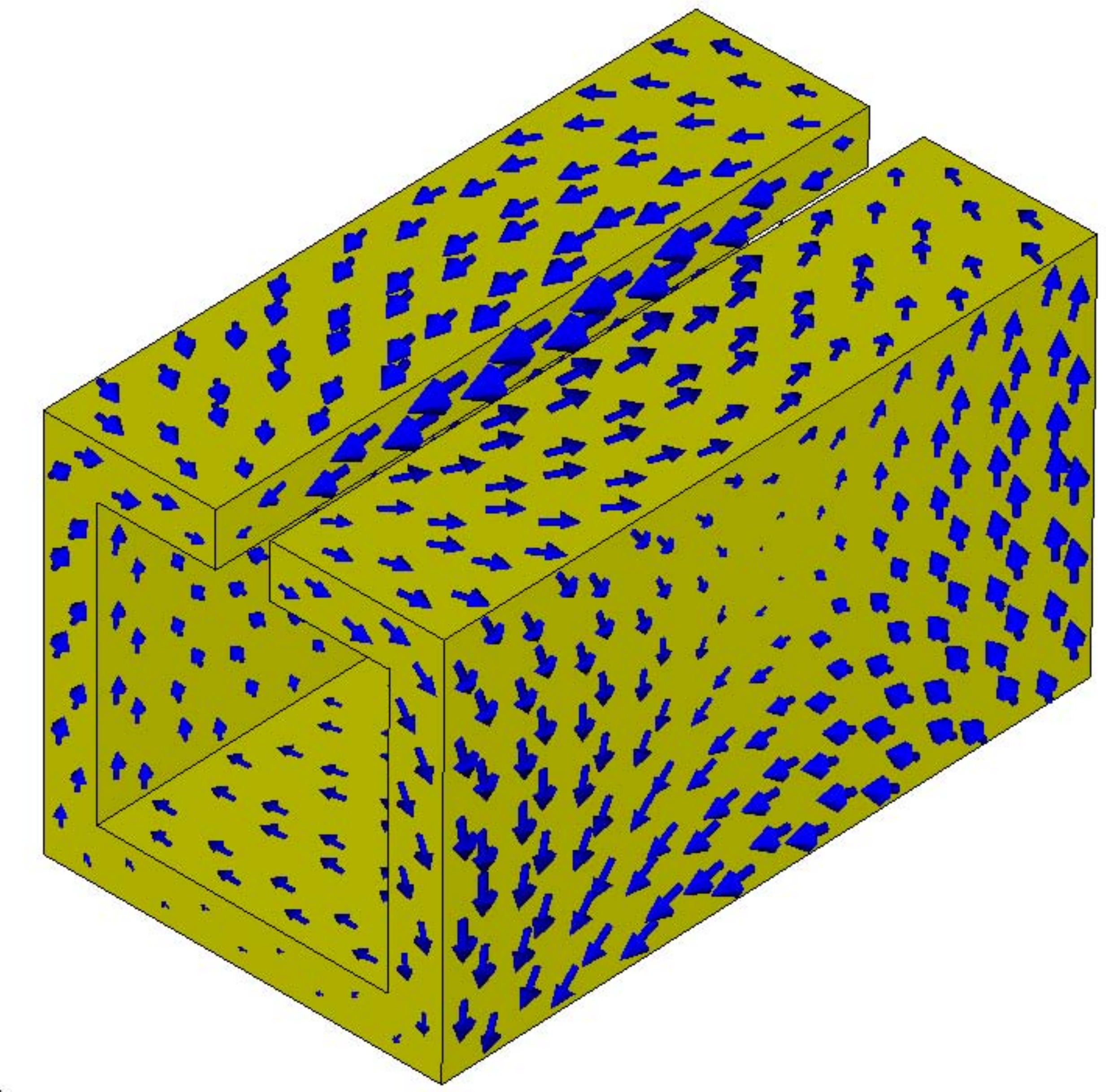}\\
\includegraphics[height=4cm]{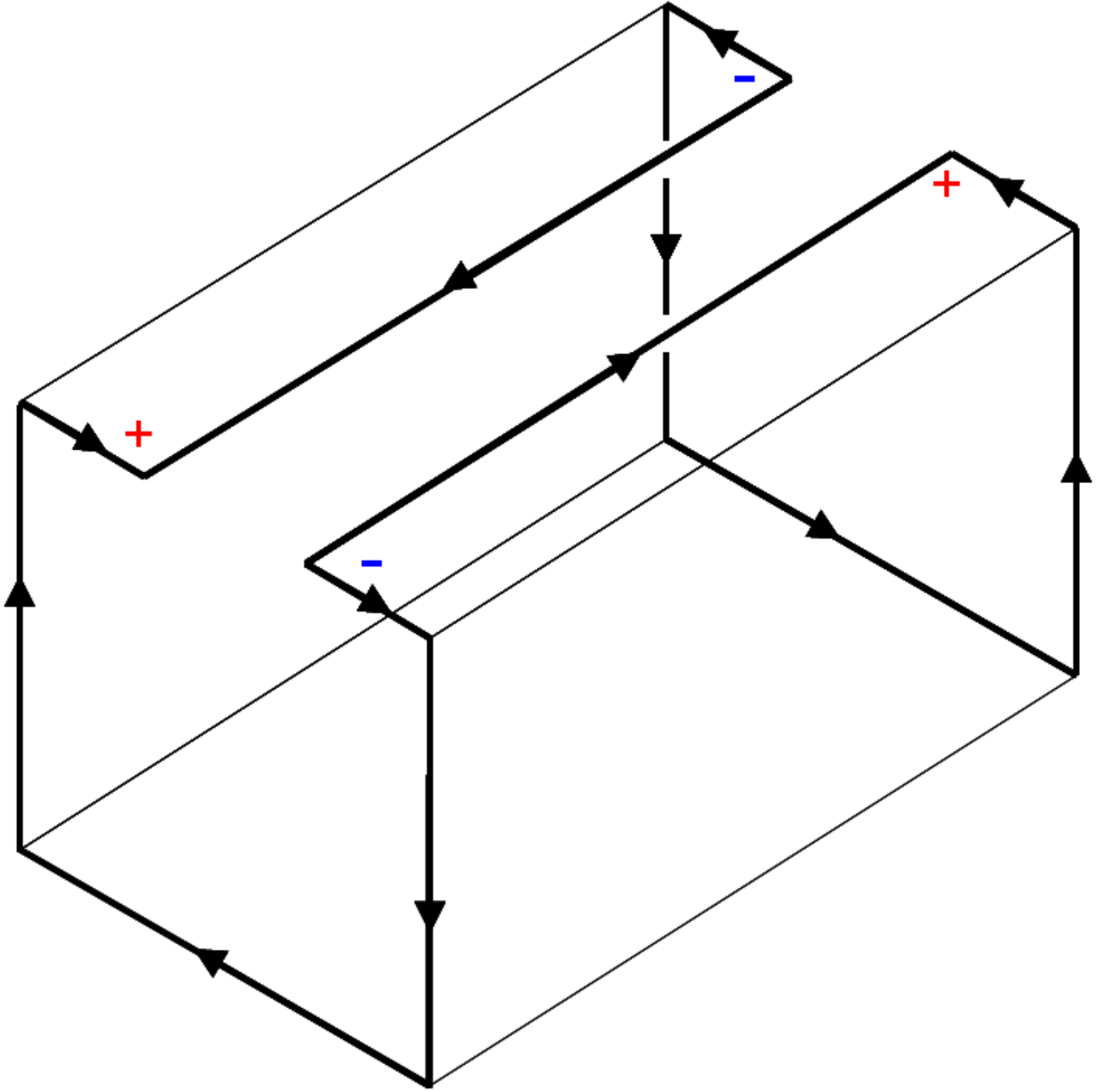}\\
\end{tabular}
\end{center}
  \vspace{-0.5cm}
\caption[currviews] {\label{currviews} Simulated currents in a 4.8\microns deep SRR in free space at a frequency of 37 THz. The schematic diagram indicates current and the areas of charge accumulation.}
\end{figure}

We explain the axial currents as being induced by the external magnetic field passing through the gap. Its time derivative creates an electromotive force in a 3-dimensional loop formed by the gap edges and the cylinder circumference at both ends. In the simulations of Figure \ref{current}, the SRRs are attached to a silicon substrate. This duplicates the experimental results well but gives rise to asymmetric current about the mid-depth plane. To see the undistorted resonant current, Figure \ref{currviews} shows simulated currents (and a schematic diagram) for free-space SRRs at the corresponding resonance at 37 THz. We observe opposing circular currents at either end of the split cylinder, with axial gap currents. The combined currents lead to charge accumulation at the gap corners, leading to electric field enhancement across the gaps. The result is thus a simultaneous concentration of electric and magnetic fields in the same region of space between the gaps. This contrasts with the case of regular LC resonances in SRRs, where the electric field is enhanced across the gaps and the induced magnetic field is normal to the plane of the SRRs.

In conclusion, our high aspect ratio PBW technique has allowed the fabrication of deep SRRs. These structures have sub-micron minimum feature size and depths of several \micron .  We observed in these Split Cylinder Resonators (SCRs) a magnetic resonance at 26 THz with distinct 3-dimensional currents. This resonance is characterized by strong axial currents and occur when the SRR depth is approximately half the wavelength of the incident radiation. Here, we have demonstrated that stretching the aspect ratio of the well-studied SRR structure can lead to the appearance of an entirely new resonant mode. Such resonances can also be expected in other metamaterial structures and will have an impact on attempts to designed 3-dimensional metamaterials for practical use. The possibility of obtaining a magnetic response from SCRs under normal incidence will create new flexibility and possibilities in the design of metamaterials.

\begin{acknowledgments}

The authors appreciate the help of Dr Chammika N B Udalagama with the software used in the Proton Beam Writing process. Dr Linus Lau of Computer Simulations Technologies provided excellent technical support. The work at the Center for Ion Beam applications was funded by the National University of Singapore grant NUS R144 000 204 646. The work done at the Singapore Synchrotron Light Source was funded by NUS Core support C-380-003-001, A*STAR/MOE RP 3979908M and A*STAR 12 105 0038 grants.

\end{acknowledgments}

%\bibliography{d:/Papers_n_ref/meta}
%\bibliographystyle{apsrev}

\end{document}